\documentclass[a4,12pt]{article}
\usepackage{graphics}
\usepackage[usenames]{color}
\usepackage[dvips]{epsfig}
\usepackage{amsmath,amssymb}
\usepackage{citesort}
\newcommand{\be}{\begin{equation}}
\newcommand{\ee}{\end{equation}}
\newcommand{\bea}{\begin{eqnarray}}
\newcommand{\eea}{\end{eqnarray}}
\newcommand{\Slash}[1]{{\ooalign{\hfil/\hfil\crcr$#1$}}}
\newcommand{\norsl}{\normalsize\sl}
\newcommand{\norsc}{\normalsize\sc}
\newcommand{\ep}{0}

\newif\ifContLineOne
\newif\ifContLineTwo
\newif\ifContLineThree
\def\conC#1{\vbox{\ialign{##\crcr
  \ifContLineThree\hrulefill\else\vphantom{\hrulefill}\fi\crcr
  \noalign{\kern3.2pt\nointerlineskip}
  \ifContLineTwo\hrulefill\else\vphantom{\hrulefill}\fi\crcr
  \noalign{\kern3.2pt\nointerlineskip}
  \ifContLineOne\hrulefill\else\vphantom{\hrulefill}\fi\crcr
  \noalign{\nointerlineskip}
  $\hfil\textstyle{\vbox to 14pt{}#1}\hfil$\crcr}}}
\def\DrawLeg#1#2{
  \kern-.2pt              % back up half width of leg
  \dimen2 =#1             % =height of whatever is underneath leg
  \advance\dimen2 by 2pt  % 2pt space below bottom of leg
  \dimen3 = 10.6pt        % base value of height of top of leg
  \dimen4 =3.6pt          % add this much time 1 2 or 3 to base value
  \advance\dimen3 by -\dimen2 
  \multiply\dimen4 by #2
  \advance\dimen3 by \dimen4
  \raise\dimen2 \hbox{\vrule height\dimen3 width .4pt} % draw it
  \kern-.2pt}             % and back up half width of line
\def\begC#1#2{\setbox0 =\hbox{$\textstyle{#2}$}
  \dimen0=.5\wd0 \dimen1=\ht0
  \conC{\hskip\dimen0}
  \count255=#1
  \ifnum\count255 =1 \ContLineOnetrue\else
  \ifnum\count255 =2 \ContLineTwotrue\else
  \ifnum\count255 =3 \ContLineThreetrue\fi\fi\fi
  \DrawLeg{\dimen1}{\count255}
  \conC{\hskip\dimen0}
  \kern-\dimen0\kern-\dimen0 \box0}
\def\endC#1#2{\setbox0 =\hbox{$\textstyle{#2}$}
  \dimen0=.5\wd0 \dimen1=\ht0
  \conC{\hskip\dimen0}
  \count255=#1
  \ifnum\count255 =1 \ContLineOnefalse\else
  \ifnum\count255 =2 \ContLineTwofalse\else
  \ifnum\count255 =3 \ContLineThreefalse\fi\fi\fi
  \DrawLeg{\dimen1}{\count255}
  \conC{\hskip\dimen0}
  \kern-\dimen0\kern-\dimen0 \box0}
\begin{document}

%% Title + Authors 
\title{Operator product expansion for B-meson 
%light-cone 
distribution amplitude
and dimension-5 HQET operators}   

\author{
{\norsc Hiroyuki Kawamura${}^{a,b}$ and Kazuhiro Tanaka${}^c$}\\
\\
\norsl ${}^a$ Radiation Laboratory, RIKEN, Wako 351-0198, Japan \\
\norsl ${}^b$ Department of Mathematical Sciences, University of Liverpool,\\
\norsl           Liverpool, L69 3BX, United Kingdom\\
\norsl ${}^c$ Department of Physics, Juntendo University, Inba, Chiba 270-1695, Japan}
\date{} 
\maketitle 
\thispagestyle{empty}

\begin{abstract}
When the bilocal heavy-quark effective theory (HQET) operator for the $B$-meson 
distribution amplitude has
a light-like distance $t$ between the quark and antiquark fields, 
the scale $\sim 1/t$ separates the UV and IR regions,
which induce the cusp singularity in 
radiative corrections and the mixing of multiparticle states 
in nonperturbative corrections, respectively.
We treat these notorious UV and IR behaviors simultaneously
using the operator product expansion,
with the local 
operators of dimension 
$d \le$ 5 and 
radiative corrections at order $\alpha_s$ 
for the corresponding Wilson coefficients.
The result is derived in the coordinate space, 
which manifests
the Wilson coefficients
with Sudakov-type double logarithms
and 
the higher-dimensional operators with additional 
gluons.
This result yields
the $B$-meson distribution amplitude for $t$ less than $\sim 1$ GeV$^{-1}$, in
terms of $\bar{\Lambda}=m_B-m_b$
and the two 
additional HQET parameters as matrix elements of dimension-5 operators.
The impact of these novel HQET parameters on the integral relevant to exclusive 
$B$ decays, $\lambda_B$, is also discussed.
\end{abstract}

\newpage

\pagestyle{plain}
\setcounter{page}{1}
\setcounter{equation}{0}
\renewcommand{\theequation}{\arabic{equation}}

\vskip 1cm
For the exclusive $B$-meson decays, such as $B \to \pi \pi$, 
$\rho \gamma, \ldots$, systematic methods have been developed
using QCD factorization based on the heavy-quark 
limit~\cite{Beneke:2000ry,Bauer:2001cu,Li:2003yj}.
Among the building blocks in the corresponding factorization formula of the decay amplitude,
essential roles are played by
the light-cone 
distribution amplitudes (LCDAs) for the participating mesons, 
which include
nonperturbative long-distance contributions.
In particular, in addition to the LCDAs for the light mesons $\pi, \rho$, etc.,
produced in the final state,
those for the $B$ meson~\cite{Szczepaniak:1990dt} also participate in 
processes where large momentum is transferred to the soft 
spectator quark via hard gluon
exchange~\cite{Beneke:2000ry,Bauer:2001cu,Li:2003yj,Beneke:2000wa,BBNL,Beneke:2005vv,Bell08}.
The leading quark-antiquark component of the
$B$-meson LCDA is defined
as the vacuum-to-meson matrix element~\cite{Grozin:1997pq}\footnote{For 
three-quark 
LCDAs for the $\Lambda_b$ baryon, see \cite{Ball:2008fw}.}:
\begin{equation}
\tilde{\phi}_+(t, \mu)
=
\frac{1}{iF(\mu)}
\langle 0|
\bar{q}(tn)
{\rm P}e^{ig\int_0^td\lambda n\cdot A(\lambda n)}
\Slash{n}
\gamma_5h_v(0) 
|\bar{B}(v)\rangle 
=\int d\omega e^{-i\omega t}
\phi_+(\omega, \mu)\ .
\label{eq1}
\end{equation}
Here,
the bilocal operator 
is built of 
the $b$-quark and light-antiquark 
fields, $h_v(0)$ and $\bar{q}(tn)$, linked by the Wilson line
at a light-like 
separation 
$tn$, with $n_\mu$ as the light-like vector 
($n^2 =0$, $n\cdot v=1$), and $v_\mu$ representing the
4-velocity of the $B$ meson;
a difference between (\ref{eq1}) and the familiar pion-LCDA~\cite{BL} is that $h_v(0)$ 
is an effective field in the heavy-quark effective theory (HQET)~\cite{Neubert:1994mb}.
$\mu$ denotes
the scale where the 
operator is renormalized, and 
$F(\mu)$ is the decay constant in HQET, $F(\mu)=-i\langle 0|
\bar{q}
\Slash{n}
\gamma_5h_v 
|\bar{B}(v)\rangle$.
The RHS in (\ref{eq1}) defines 
the momentum representation, 
with $\omega v^+$ denoting the LC component of the momentum of the light antiquark.

The ``IR structure'' of the $B$-meson LCDA was studied~\cite{KKQT}
using constraints from the equations of motion (EOM), 
$\bar{q}\overleftarrow{\Slash{D}}  = v \cdot \overrightarrow{D} h_v  = 0$,
and heavy-quark symmetry, $\Slash{v}h_v=h_v$. These constraints are solved to give (\ref{eq1}) as
$\phi_+ (\omega ) = \phi_+^{(WW)} (\omega ) + \phi_+^{(g)} (\omega )$, where
the first term is expressed by the analytic formula 
$\phi_+^{(WW)} (\omega ) = \omega \theta (2\bar{\Lambda}  - \omega )/( 2\bar{\Lambda}^2 )$
in terms of a HQET parameter~\cite{Neubert:1994mb}, 
$\bar{\Lambda}=m_B -m_b$, which represents the mass difference
between the $B$-meson and $b$-quark. The second term, $\phi _+^{(g)}$, is obtained as a certain 
integral
of the matrix element of 
the three-body LC operator,
$\langle \bar{q}(tn)G_{\alpha \beta}(un)\sigma_{\lambda \eta} h_v (0) \rangle$,
where\footnote{For simplicity the Wilson lines connecting the constituent fields are suppressed, 
and $\langle \cdots \rangle \equiv \langle 0| \cdots|\bar{B}(v) \rangle$.}
the nonperturbative gluons participate as the field strength tensor $G_{\alpha \beta}$; 
see (\ref{ide}) below 
for the integro-differential equation, which is derived from the 
relevant EOM constraints and gives 
$\phi_+ (\omega ) = \phi_+^{(WW)} (\omega ) + \phi_+^{(g)} (\omega )$ as its solution 
(see also \cite{Huang:2005kk}).
In addition, the ``UV structure'' of the $B$-meson 
DA was studied~\cite{Lange:2003ff}
by calculating the 1-loop renormalization of the bilocal operator in (\ref{eq1}).
The ``vertex-type'' correction around a ``cusp'' between the two Wilson lines,
the light-like Wilson line of (\ref{eq1}) and the time-like Wilson line from 
$h_v(0)={\rm P}\exp [ig\int_{-\infty}^0d\lambda v\cdot A(\lambda v)]h_v(-\infty v)$,
produces the ``radiation tail'', given 
by $\phi_+(\omega, \mu) \sim -C_F \alpha_s \ln(\omega/\mu)/(\pi \omega)$
for $\omega \gg \mu$, where $C_F = (N_c^2-1)/(2N_c)$.
This implies that the moments $\int_0^\infty  d\omega  \omega^j \phi _+ (\omega, \mu)$,
which would correspond to 
matrix elements of 
the local operators 
$\bar{q}(0)(n\cdot D )^j
\Slash{n}
\gamma_5 h_v (0)$,
are divergent,
reflecting 
``cusp singularity''~\cite{Lange:2003ff,Korchemsky:1987wg}.

However, these results do not represent the whole story.
As we shall demonstrate explicitly,
a full description of (\ref{eq1}) actually involves a complicated mixture 
of the IR and UV structures;
e.g., 
the above-mentioned
functional forms $\phi_+^{(WW)} (\omega )$, $\phi_+^{(g)} (\omega )$
from the EOM constraints are subject to additional effects from radiative corrections 
when combined with the UV structure 
and are profoundly modified, in particular, for large values of $\omega$
(see the discussion below (\ref{ide})).
To investigate the behavior 
of the $B$-meson DA 
incorporating both IR and UV 
structures, 
we first calculate the radiative corrections, 
taking into account hard and soft/collinear loops.
The one-particle-irreducible 1-loop diagrams (1LDs) for the 2-point function
$\langle \bar{q}(tn)\Slash{n}\gamma_5h_v(0) \rangle$ 
of (\ref{eq1}) yield\footnotemark[2] 
\begin{eqnarray}
&&
\!\!\!\!\!\!
{\rm 1LDs}=
\frac{\alpha_s C_F }{2\pi }
\int_0^1 {d\xi } \left[ \left\{  - \left( \frac{1}{2\varepsilon_{UV}^2} 
+ \frac{L}{\varepsilon_{UV}} + L^2+ \frac{5\pi^2 }{24} \right)
\delta (1 - \xi ) 
\right. \right.
+ 
\left( \frac{1}{\varepsilon_{UV}} - \frac{1}{\varepsilon_{IR}} \right)
\left( \frac{\xi}{1 - \xi} \right)_+
\nonumber\\   
&&
\!\!\!- \left.   
\left( \frac{1}{2\varepsilon_{IR} } 
+ L \right)\! \right\}\!
\langle 
\bar{q}(\xi tn) \Slash{n}
\gamma_5 h_v (0) 
\rangle  
-
\left.
   t\left( \frac{1}{\varepsilon_{IR}} + 2 L- 1 - \xi  \right)\!
\langle 
\bar{q}(\xi tn)v \cdot \overleftarrow{D}  \Slash{n}
\gamma_5 h_v (0)
\rangle  
\right] \!\!+\! \cdots,
\label{eq2}
\end{eqnarray}
in $D=4-2\varepsilon$ dimensions and Feynman gauge,
where 
$L\equiv \ln\left[i(t-i0) \mu e^{\gamma_E}\right]$
with the ${\overline{\rm MS}}$ scale $\mu$ and the Euler constant $\gamma_E$, 
and the ``$-i0$'' prescription comes from the position of
the pole in the relevant propagators 
in the coordinate-space 
(see (\ref{loopint}) below).
The ``vertex-type'' correction 
that connects the light-like Wilson line and the $\bar{q}(tn)$ field
in (\ref{eq1}) 
is associated with only the massless degrees of freedom; thus, the correction
yields the term with the ``canceling'' UV and IR poles, $1/\varepsilon_{UV}-1/\varepsilon_{IR}$,
from the scaleless loop-integral, 
and with
the ``plus''-distribution $(\xi/(1-\xi) )_+$ as the splitting function.
This term 
is identical to
the corresponding correction for 
the case of 
the pion LCDA, where the contributions from all 1-loop 
diagrams have the same $1/\varepsilon_{UV}-1/\varepsilon_{IR}$ structure 
%characteristic of a massless case 
%accompanying 
with the Brodsky-Lepage kernel~\cite{BL} as the total splitting function.
However, the other terms in (\ref{eq2})
have 
``non-canceling'' UV and IR poles:
another vertex-type correction, connecting the light-like Wilson line 
and the $h_v(0)$ field
in (\ref{eq1}),
gives~\cite{Lange:2003ff} the 
terms proportional to 
$\delta (1-\xi)$, which contain
the double as well as single UV pole, corresponding to the cusp singularity mentioned above.
The ``ladder-type'' correction, connecting the two quark fields in (\ref{eq1}), gives
all the remaining terms in (\ref{eq2}), which contain
the IR poles and are associated with not only the bilocal operator in (\ref{eq1}), 
but also the higher dimensional operators (see the discussion below (\ref{loopint}));
note that the ellipses in (\ref{eq2}) are expressed by the operators
involving 
two or more additional
covariant derivatives.

The renormalized LCDA is obtained by subtracting the UV poles from (\ref{eq2}) 
with the trivial quark self-energy corrections complemented.
Here, the term with the plus-distribution $(\xi/(1-\xi) )_+$ 
is analytic (Taylor expandable) at $t=0$, similar to the pion LCDA, but
the other terms
are not
analytic
due to the presence of logarithms $L$, $L^2$~\cite{Lange:2003ff,Braun:2003wx}.
In particular, the nontrivial dependence of the latter terms on $t\mu$ through $L$
implies that the scale $\sim 1/t$ separates the UV and IR regions.
Thus, we have to use the operator product expansion (OPE) to
treat the different UV and IR behaviors simultaneously:
the coefficient functions absorb all the singular logarithms,
while, for the local operators to absorb the IR poles, we have to take into account many 
higher dimensional operators.
Such OPE with local operators is useful when the separation $t$
is less than the typical distance scale of quantum fluctuation,
i.e., when $t\lesssim 1/\mu$.
The OPE result can be 
evolved 
from a low $\mu$ 
to a higher scale
using the Brodsky-Lepage-type kernel and the Sudakov-type operator 
with the cusp anomalous dimension~\cite{Lange:2003ff,Braun:2003wx}
associated with the single-UV-pole terms in (\ref{eq2}).

An OPE for the $B$-meson LCDA (\ref{eq1}) 
was discussed 
in \cite{Lee:2005gza},
taking into account the local operators of dimension 
$d \le$ 4
and 
the NLO ($O(\alpha_s)$)
corrections to the corresponding Wilson coefficients 
in a 
``cutoff scheme'', where
an additional momentum cutoff $\Lambda_{UV}$ ($\gg \Lambda_{\rm QCD}$) 
was introduced, 
and the OPE, in powers of $1/\Lambda_{UV}$, was derived for the regularized moments,
$\int_0^{\Lambda_{UV}} d\omega  \omega^j 
\phi_+ (\omega, \mu)$,
in particular, for the first two moments with $j=0, 1$.
In this Letter, we derive the OPE for 
(\ref{eq1}),
taking into account the local operators of dimension 
$d \le$ 5
and calculating 
the corresponding Wilson coefficients
at NLO accuracy.
Following the discussion above, we carry out the calculation for $t\lesssim 1/\mu$
in the coordinate space and
in the $\overline{\rm MS}$ scheme, 
so that there is no need to introduce any additional cutoff.

The most complicated part of this calculation 
is the reorganization of contributions from 
(many) Feynman diagrams 
in terms of the
matrix element
of 
gauge-invariant operators, including higher dimensional operators.
In particular, to derive the Wilson coefficients associated with the dimension-5
operators such as $\bar{q}G_{\alpha\beta}\Slash{n}  \gamma_5 h_v$, 
we have to compute the 1-loop diagrams for the 3-point function,
as well as those for the 2-point function as in (\ref{eq2}),
where the former diagrams are obtained by attaching the external gluon line
to the latter diagrams in all possible ways. 
To perform the calculation in a systematic and economic way, we employ the background 
field method~\cite{Abbott:1980hw}.
We decompose the quark and gluon fields into the ``quantum'' and ``classical'' parts,
where the latter part represents the nonperturbative long-distance 
degrees of freedom inside the $B$ meson as a background field and satisfies the classical EOM.
The quark and gluon propagators for the quantum part contain 
their coupling with an arbitrary number of background fields, and 
each building block of the Feynman diagrams obeys the exact 
transformation property under the gauge transformation for the background fields.
We use the Fock-Schwinger gauge, $x^\mu  A_\mu^{(c)} (x)=0$,
for the background gluon field $A_\mu^{(c)}$.
This gauge condition is solved to give 
$A_\mu^{(c)}(x) = \int_0^1 {du} u x^\beta  G_{\beta \mu}^{(c)} (ux)$~\cite{Abbott:1980hw},
which allows us to reexpress each Feynman diagram in terms of the
matrix element of the operators associated with the field strength tensor.
Using this framework, 
the tree-level matching to derive our OPE
can be performed replacing each constituent
field in 
(\ref{eq1}) 
with the corresponding background
field.
The {\it classical} bilocal operator can be Taylor expanded, 
and
we obtain the OPE at the tree level
with the local operators of dimension-3, -4, and -5
(see the $O(\alpha_s^0)$ terms in (\ref{eq3}) below).

For the 1-loop matching, we calculate the 1-loop corrections to 
the 2- and 3-point functions with the insertion of the bilocal operator in (\ref{eq1}),
taking into account the mixing of operators of dimension 
$d \le$ 5.
Apparently, the mixing through the UV region of the loop momenta 
arises only in the 2-point function,
and the result can be immediately obtained from (\ref{eq2}); however, additional
mixing can arise
in both 2- and 3-point functions accompanying the IR poles.
We perform the loop integration in the coordinate space 
using the Schwinger (``heat kernel'') representation 
of the Feynman amplitudes under the background fields~\cite{Abbott:1980hw}.
For the calculation of the 3-point function,
the Fock-Schwinger 
gauge ensures that the Wilson line in (\ref{eq1}), 
as well as the heavy-quark field,
does not couple directly to the background gluons,
while a massless quark or gluon line couples to them.
For example, the ``ladder-type'' correction diagrams for the 2- and 3-point functions
are obtained by connecting the two quark fields in (\ref{eq1}) 
using the gluon propagator 
in the 
background gluon fields~\cite{Abbott:1980hw}, and, after some manipulations
using the heat-kernel expansion, 
we encounter the (coordinate-space) loop integrations
of the type
\begin{equation}
\int_0^\infty \!\!\!\! dr
\frac{\left[\frac{e^{\gamma_{E}}\mu^2}{4}\right]^{\frac{4-D}{2}}\!\!\!
\Gamma\left(\frac{D}{2}-m\right)
r^{1-m+n}}{(-r^2-2tr+i\ep)^{\frac{D}{2}-m}}
= \frac{(-1)^{n+1} \Gamma(n)(2t)^{m+n-2}}{\Gamma (m+n-1)} \!
\left[\frac{1}{2\varepsilon_{IR}}
+L+ \frac{1}{2}S_{n-1}-S_{m+n-2} \right]\! ,
\label{loopint}
\end{equation}
in $D$ ($=4-2\varepsilon$) dimensions, with $n, m = 1,2,\ldots$, and $S_n = \sum_{k=1}^n (1/k)$.
Here, the integral over the proper time $r$, associated with the propagation of
the heavy quark in the corresponding diagrams,
is UV finite but IR divergent 
as $D\rightarrow 4$, yielding the RHS.
The 2-point function receives the contribution of 
the dimension-3 operator, $\bar{q}\Slash{n}\gamma_5h_v$,
accompanying 
(\ref{loopint}) with $n=m=1$ as its coefficient.
The higher dimensional operators with additional covariant derivatives 
and/or the gluon field strength tensor contribute to 2- as well as 3-point functions,
accompanying (\ref{loopint}) with $n+m \ge 2$.
Those latter contributions manifest the mixing of operators of dimension-4 and higher,
accompanying the IR poles.
Note that
this type of
mixing mechanism for ``higher twist'' operators through loop effects is forbidden
if the corresponding loop subdiagram is composed of massless degrees of freedom only.

After working out the loop integral of all 1-loop diagrams 
for the 2- and 3-point functions, 
we reorganize the result in terms of a complete set of gauge-invariant operators
of dimension $d \le 5$,
using some extension of the technique of \cite{Grozin:1997pq,KKQT}
based on the EOM and heavy-quark symmetry.
We subtract the UV poles to renormalize the bilocal operator of (\ref{eq1})
and also renormalize the local operators to absorb the IR poles.
(The details of the above-mentioned procedures to treat the 2- and 3-point functions at 1-loop 
will be presented elsewhere~\cite{KT08}.)
Combining 
the result 
with the tree-level result obtained above,
we obtain the OPE for the bilocal 
operator in (\ref{eq1}),  
$\bar{q}(tn)
{\rm P}e^{ig\int_0^td\lambda n\cdot A(\lambda n)}
\Slash{n}\gamma_5h_v(0)
=\sum_{i}C_i(t,\mu) {\cal O}_i(\mu)$,
to the desired accuracy as
\begin{eqnarray}
&&
\!\!\!\!\!\!
\bar{q}(tn)
{\rm P}e^{ig\int_0^td\lambda n\cdot A(\lambda n)}
\Slash{n}\gamma_5h_v(0)
=
\left[1-  \frac{\alpha_sC_F}{4\pi}
\left(2L^2+2L+ \frac{5\pi^2}{12}\right)\right]{\cal O}^{(3)}_1
\nonumber\\&&
\;\;\;\;\;\;\;\;\;\;\;\;\;\;\;\;\;\;\;\;\;\;
-it\left\{
\left[
1- \frac{\alpha_sC_F}{4\pi}
\left(2L^2+L+\frac{5\pi^2}{12}\right)
\right] {\cal O}^{(4)}_1
-\frac{\alpha_sC_F}{4\pi}\left(4L-3\right){\cal O}^{(4)}_2
\right\}
\nonumber\\&&
\;\;\;\;\;\;\;\;\;\;\;
-\frac{t^2}{2}\left\{
\left[1- \frac{\alpha_sC_F}{4\pi}
\left(
2L^2+\frac{2}{3}L+\frac{5 \pi^2}{12}
\right)\right]{\cal O}^{(5)}_1
\right.
-\frac{\alpha_sC_F}{4\pi}
\left(4L-\frac{10}{3}\right)\left( {\cal O}^{(5)}_2
+ {\cal O}^{(5)}_3 \right)
\nonumber\\&&
\;\;\;\;\;\;\;\;\;\;\;
+ 
\frac{\alpha_s}{4\pi}
\left[ C_F\left(-4L+\frac{10}{3}\right)
+ C_G\left(7L-\frac{13}{2}\right)
\right]{\cal O}^{(5)}_4
+ 
\frac{\alpha_s}{4\pi} \left(-\frac{4}{3}C_F +C_G \right) \left(L-1\right)
{\cal O}^{(5)}_5
\nonumber\\&&
\;\;\;\;\;\;\;\;\;\;\;\;
+
\frac{\alpha_s}{4\pi} \left(-\frac{2}{3}C_F +C_G \right) \left(L-1\right)
{\cal O}^{(5)}_6
\left.
+ 
\frac{\alpha_s}{4\pi}\left(-\frac{1}{3}C_F +\frac{1}{4}C_G \right) \left(L-1\right)
{\cal O}^{(5)}_7
\right\}\ .
\label{eq3}
\end{eqnarray}
Here and below, 
$C_G=N_c$, $\mu$ is the $\overline{\rm MS}$ scale, and $\alpha_s \equiv \alpha_s(\mu)$.
A basis of local operators of dimension-$d$, ${\cal O}^{(d)}_k$
($k=1,2,\ldots$), 
is defined as
\begin{equation}
{\cal O}^{(3)}_1 = \bar{q}\Slash{n}\gamma_5h_v\ , \;\;\;\;\;\;
\left\{ {\cal O}^{(4)}_k \right\} = \left\{ \bar{q}(i n\cdot \overleftarrow{D})
\Slash{n}\gamma_5h_v\ , \;\;
\bar{q}(iv\cdot \overleftarrow{D})
\Slash{n}\gamma_5h_v \right\}\ , 
\label{basis34}
\end{equation}
\bea
\left\{ {\cal O}^{(5)}_k  \right\}&& = 
 \left\{ \bar{q}(in\cdot \overleftarrow{D})^2
\Slash{n} \gamma_5 h_v\ ,  \;\; 
\bar{q}(iv\cdot \overleftarrow{D})
(in\cdot \overleftarrow{D})\Slash{n}\gamma_5 h_v\ , \;\;
\bar{q}(iv\cdot \overleftarrow{D})^2
\Slash{n}\gamma_5 h_v\ ,
\right.
\nonumber\\
&&\left. 
\!\!\!\!\! 
\!\!\!\!\!
\bar{q}igG_{\alpha\beta}v^\alpha n^\beta \Slash{n}
\gamma_5h_v\ ,  \;\;
\bar{q}igG_{\alpha\beta}\gamma^{\alpha}n^{\beta}
\Slash{\bar{n}}\gamma_5h_v\ , \;\;
\bar{q}igG_{\alpha\beta}\gamma^\alpha v^\beta
\Slash{\bar{n}}
\gamma_5h_v\ , \;\;
\bar{q}gG_{\alpha\beta}\sigma^{\alpha\beta}
\Slash{n}\gamma_5h_v \right\}\ , 
\label{basis5}
\eea
with another light-like vector, $\bar{n}^2=0$,
as $v_\mu=(n_\mu+\bar{n}_\mu)/2$. 
The double logarithm $L^2$ in the coefficient functions
originates from the cusp singularity (see (\ref{eq2})).
The 1-loop corrections for the 3-point function 
induce only 
${\cal O}^{(5)}_{4,5,6,7}$
associated with the field-strength tensor, 
while those for the 2-point function
induce all ten operators 
of (\ref{basis34}), (\ref{basis5}) 
through the use of the EOM.

Taking the matrix element $\langle \cdots \rangle \equiv \langle 0| \cdots|\bar{B}(v) \rangle$
of (\ref{eq3}), we can derive the OPE form of 
the $B$-meson LCDA (\ref{eq1}).
The matrix elements of the local operators in (\ref{basis34}), (\ref{basis5}) 
are known to be related to a few nonperturbative parameters
in the HQET, using the EOM and heavy-quark symmetry as demonstrated in
\cite{Grozin:1997pq,KKQT}:
$\langle {\cal O}^{(4)}_1 
\rangle= 4iF(\mu)  \bar{\Lambda}/3$, 
$\langle {\cal O}^{(4)}_2 
\rangle
= iF(\mu) \bar{\Lambda}$, 
where $F$ and $\bar{\Lambda}$ were introduced below (\ref{eq1}),
and all seven matrix elements $\langle {\cal O}^{(5)}_k  \rangle$ for the dimension-5 
operators (\ref{basis5})
can be expressed by $F$,
$\bar{\Lambda}$ and two additional HQET parameters $\lambda_E$ and $\lambda_H$,
which are associated with the
chromoelectric and chromomagnetic fields inside the $B$ meson as
$\langle \bar{q}g\boldsymbol{E}\cdot \boldsymbol{\alpha}
\gamma_5 h_v \rangle
=F(\mu)\lambda_E^2(\mu)$ and
$\langle \bar{q}g\boldsymbol{H}\cdot \boldsymbol{\sigma}
\gamma_5h_v \rangle
=i F(\mu)
\lambda_H^2(\mu)$, respectively, in the rest frame where $v=(1,{\bf 0})$.
As a result, we obtain the OPE form for the LCDA (\ref{eq1}),
\begin{eqnarray}
&&\!\!\!\!
\tilde{\phi}_+(t,\mu)
=
1- \frac{\alpha_s C_F}{4\pi}
\left(2L^2+2L+\frac{5 \pi^2}{12}\right)
-it\frac{4\bar{\Lambda}}{3} 
\left[1- \frac{\alpha_s C_F}{4\pi}
\left(2L^2+4L-\frac{9}{4}+\frac{5\pi^2}{12} \right)
\right]
\nonumber\\
&&\!\!\!\!
-t^2 \bar{\Lambda}^2 \!\! \left[
1\! - \! \frac{\alpha_sC_F}{4\pi}\!\!
\left(2L^2+\frac{16}{3}L-\frac{35}{9}
+\frac{5\pi^2}{12} \right)\!
\right]
\!\!- \! \frac{t^2\lambda_E^2(\mu)}{3}\!
\left[1\! - \! \frac{\alpha_sC_F}{4\pi}\!
\left( \! 2L^2+2L-\frac{2}{3}
+\frac{5\pi^2}{12} \! \right)
\right.
\nonumber\\
&&\;\;
\left.
+ 
\frac{\alpha_sC_G}{4\pi}
\left(\frac{3}{4}L-\frac{1}{2}\right)
\right]
-\frac{t^2\lambda_H^2(\mu)}{6} 
\left[1- \frac{\alpha_sC_F}{4\pi}
\left(2L^2+\frac{2}{3}
+\frac{5\pi^2}{12} \right)
-\frac{\alpha_s C_G}{8\pi}
\left(L-1\right)
\right]\ ,
\label{eq4}
\end{eqnarray}
which takes into account the Wilson coefficients 
to $O(\alpha_s)$ and a complete set of the local operators of dimension 
$d \le$ 5;
the cusp singularity in the former leads to the double logarithms $L^2$,
while
the constraints on matrix elements of the latter from the EOM and heavy-quark symmetry
allow us to represent the result 
completely in terms of only three HQET parameters,
$\bar{\Lambda}$,
$\lambda_E$ and $\lambda_H$.
Thus, (\ref{eq4}) ``merges'' the UV~\cite{Lange:2003ff} and IR structures~\cite{KKQT} 
peculiar to the $B$-meson LCDA, 
so that it embodies novel behaviors that are completely different from those
of the pion LCDA:
$\mu$ and $t$ are strongly correlated due to the
logarithmic contributions, $L= \ln\left[i(t-i0) \mu e^{\gamma_E}\right]$,
from radiative corrections,
so that the DA is not Taylor expandable about $t=0$.
The DA receives the contributions from (many) higher dimensional operators,
in particular, from those associated with the long-distance gluon fields inside
the $B$-meson.

Fourier transforming 
(\ref{eq4}), 
we obtain the momentum 
representation,
$\phi_+(\omega, \mu)$ in (\ref{eq1}), and
can also
evaluate the regularized-moments,
$M_{j} = \int_0^{\Lambda_{UV}} d\omega  \omega^j 
\phi_+ (\omega, \mu)$,
with the 
cutoff $\Lambda_{UV}$ ($\gg  \Lambda_{\rm QCD}$);
$M_j$ corresponds to the ``regularized coefficients''
of the $t^j$ term in the Taylor expansion of the DA (\ref{eq1})
and thus 
represents the $t^j M_j=O\left( M_j /\Lambda_{UV}^j \right)$ effects on $\tilde{\phi}_+(t,\mu)$
for $t\lesssim 1/\Lambda_{UV}$.
Here, without going into the detail~\cite{KT08},
we note that, for the first two moments $M_0$ and $M_1$, 
the contributions from the first line in (\ref{eq4}),
associated with the matrix elements of
the dimension-3 and -4 operators (\ref{basis34}),
coincide with the result obtained in \cite{Lee:2005gza}, while
the second and third lines\footnote{
They also 
allow us to derive 
$M_2$,
up to the corrections 
that are suppressed for $\Lambda_{UV}\gg \Lambda_{\rm QCD}$~\cite{KT08};
from dimension counting~\cite{Lee:2005gza},
the local operators 
of dimension $d$ ($\ge 3$) contribute to $M_j$ as 
$\sim \Lambda_{UV}^j (\Lambda_{\rm QCD}/\Lambda_{UV})^{d-3}$.
}
in (\ref{eq4}), which are generated from the dimension-5 operators (\ref{basis5}), 
yield the new power-correction terms down by $\Lambda_{\rm QCD}/\Lambda_{UV}$,
as controlled by the corresponding Wilson coefficients.
A similar pattern is observed for the asymptotic behavior 
of $\phi_+(\omega, \mu)$ for $\omega \gg \Lambda_{\rm QCD}$.
In view of our coordinate-space OPE approach, 
the UV divergence in the moments~\cite{Grozin:1997pq,Lange:2003ff,Lee:2005gza},
$M_j \rightarrow \infty$ as $\Lambda_{UV} \rightarrow \infty$, 
reflects the non-analyticity of (\ref{eq3}) and (\ref{eq4}) at $t=0$
due to the presence of logarithms $L$ and $L^2$ in the Wilson coefficients
(see also the discussion in \cite{Braun:2003wx}).

Our OPE results (\ref{eq3}) and (\ref{eq4})
give a model-independent description 
of the $B$-meson LCDA 
when $t \lesssim 1/\mu$ ($\leq 1/\Lambda_{\rm QCD}$),
taking into account the UV and IR structures simultaneously.
It is instructive to draw a comparison 
with the previous results mentioned below (\ref{eq1}), concerning
UV~\cite{Lange:2003ff} or IR structure~\cite{KKQT}.
In \cite{Lange:2003ff}, the renormalization group (RG) evolution of (\ref{eq1})
was derived. The corresponding evolution kernel is determined
by the (single) UV poles in (\ref{eq2}) as the sum of 
the cusp anomalous dimension and the plus-distribution terms.
One can prove that our LCDA (\ref{eq4})
satisfies the RG equation of \cite{Lange:2003ff}, 
taking into account $d \bar{\Lambda}/d\mu=0$ and~\cite{GN97}
\begin{equation}
\mu \frac{d}{d\mu}
\left( 
\begin{array}{c}
\lambda_E^2(\mu)\\
\lambda_H^2(\mu)
\end{array}
\right)
= -\frac{\alpha_s}{4\pi}
\left( 
\begin{array}{cc}
\frac{8}{3} C_F+\frac{3}{2}C_G &\;\;\;\;\; \frac{4}{3}C_F  -\frac{3}{2}C_G \\
\frac{4}{3}C_F  -\frac{3}{2}C_G &\;\;\;\;\; \frac{8}{3}C_F  +\frac{5}{2}C_G 
\end{array}
\right)
\left( 
\begin{array}{c}
\lambda_E^2(\mu)\\
\lambda_H^2(\mu)
\end{array}
\right)\ .
\label{elam}
\end{equation}
Thus, our results, (\ref{eq3}) and (\ref{eq4}) via the 1-loop matching, are completely 
consistent with the UV structure obtained at the 1-loop level in \cite{Lange:2003ff}. 
As a result, the scale dependence of the $B$-meson LCDA (\ref{eq1}) 
with $t \lesssim 1/\Lambda_{\rm QCD}$ is represented by the two-step evolution:
it is governed by the solution of the evolution equation 
in terms of the Sudakov-type and 
Brodsky-Lepage-type scale dependence from the high scale $\mu_i\simeq\sqrt{m_b \Lambda_{\rm QCD}}$,
associated with the QCD factorization formula, to the scale $\mu \lesssim 1/t$,
while that for the lower scale is governed by the anomalous dimensions 
of local operators like (\ref{elam}). Here, in principle, we can take advantage of
the RG improvement as usual.
To achieve the 
control at the next-to-leading logarithmic
accuracy,
we take into account the 2-loop 
as well as 1-loop cusp anomalous dimension~\cite{Korchemsky:1987wg}
for the Sudakov-type evolution 
from 
$\mu_i$ to $\mu$ 
(see, e.g., \cite{BBNL});
similarly, the evolution for the lower scale requires 
the 2-loop anomalous dimensions of $\lambda_E^2$ and $\lambda_H^2$,
but they are unknown at present.

On the other hand, the connection of the present result (\ref{eq4}) with the IR structure
of \cite{KKQT} is not straightforward: 
from the EOM and heavy-quark symmetry constraints, 
a set of relations between the two- and three-particle LCDAs was 
obtained in \cite{KKQT}.  
Using these relations, we can derive
the following integro-differential equation for the LCDA (\ref{eq1}),
\bea
t\frac{d\tilde{\phi}_{+}(t)}{dt} &&
\!\!
+\left(2i \bar{\Lambda}t +1\right)\tilde{\phi}_{+}(t)-\frac{1}{t}\int_{0}^{t}dt'
\tilde{\phi}_{+}(t')
+\left.
2t^2 \int_0^1 du
           \right\{ \left(u+1\right) \tilde{\Psi}_A (t,u) 
\nonumber\\
&&+\;
u \tilde{\Psi}_V (t,u) 
+\left.  \tilde{X}_A (t,u) 
-\frac{u}{t^3}\int_{0}^{t}dt' {t'}^2 
           \left[ \tilde{\Psi}_A (t',u) - \tilde{\Psi}_V (t',u) \right]\right\}
 =0\ ,
\label{ide} 
\eea
where $\tilde{\Psi}_A(t,u)$, $\tilde{\Psi}_V (t,u)$, and
$\tilde{X}_A (t,u)$ are the three-particle LCDAs
introduced in \cite{KKQT} (see also \cite{Huang:2005kk}):
when $t \lesssim 1/\Lambda_{\rm QCD}$, 
we substitute (\ref{eq4}) for $\tilde{\phi}_+$,
while the three-particle LCDAs are given as
$\tilde{\Psi}_A= \frac{1}{3}\lambda_E^2$, 
$\tilde{\Psi}_V=\frac{1}{3}\lambda_H^2$, and
$\tilde{X}_A=0$
in terms of matrix elements of local operators of dimension 
$d\le 5$~\cite{KKQT}, omitting the contributions from
higher dimensional operators as well as from radiative corrections.
Then the $O(\alpha_s^0)$ contribution to the LHS of (\ref{ide})
vanishes up to the corrections of $O\left((t\Lambda_{\rm QCD})^3 \right)$
which are beyond the present accuracy, but
the 
$O(\alpha_s)$ contribution
proves to yield a nonzero result as
$(\alpha_s C_F /4 \pi)\left[
-8 L + it \bar{\Lambda}  \left( 32L/3+ 3/2 \right) +\cdots \right]$.
Thus, (\ref{ide}) receives corrections at order $\alpha_s$;
namely, 
the relations of \cite{KKQT} 
are satisfied 
by the nonperturbative matrix elements of local operators in the OPE (\ref{eq3}),
but the $O(\alpha_s)$ loop effects in the corresponding Wilson coefficients
prevent those relations from being exact 
for 
the DA (\ref{eq4}) 
(see also the discussion in \cite{Braun:2003wx,Bell:2008er}).
Such a violation of relations of the type (\ref{ide}) at order $\alpha_s$
in perturbation theory
is peculiar to the heavy-meson LCDAs in the HQET and does not arise 
for the case of the (higher twist) LCDAs 
for the light mesons, $\pi, \rho$, etc.~\cite{Braun:1990iv}.

Our OPE form (\ref{eq4}) allows us to parameterize all nonperturbative contributions 
in the $B$-meson LCDA (\ref{eq1}) for
$t \lesssim 1/\Lambda_{\rm QCD}$
by a few HQET parameters.
Here, we evaluate (\ref{eq4}) at the scale $\mu =1$~GeV, 
using the information available for 
those HQET parameters.
First of all, we note that $\bar{\Lambda}=m_B - m_b$ in (\ref{eq4}) is defined by
the $b$-quark pole mass $m_b$. 
Following \cite{Lee:2005gza}, 
we eliminate $\bar{\Lambda}$ in favor of a short-distance parameter, $\bar{\Lambda}_{DA}$,
free from IR renormalon ambiguities~\cite{Bigi:1994} and written as
$\bar{\Lambda}
   = \bar{\Lambda}_{DA}(\mu)
    \left[ 1 + (7/16\pi) C_F\alpha_s
\right] 
-( 9/8\pi)\mu C_F\alpha_s$,
to one-loop accuracy;
$\bar{\Lambda}_{DA}(\mu)$ can be related to another short-distance mass parameter
whose value is extracted from analysis of the spectra in inclusive decays $B\to X_s\gamma$ 
and $B\to X_u l\,\nu$, leading to
$\bar{\Lambda}_{DA}(\mu =1~{\rm GeV})\simeq 0.52$~GeV~\cite{Lee:2005gza}. 
For the novel 
parameters associated with the
dimension-5 operators, we use the central values of
\begin{equation}
\lambda_E^2 (\mu) = 0.11 \pm 0.06~{\rm GeV}^2\ , \;\;\;\; \;\;\;\;\;\;\;\;
\lambda_H^2 (\mu) = 0.18 \pm 0.07~{\rm GeV}^2\ ,
\label{lambdaEH}
\end{equation}
at $\mu =1$~GeV,
which were 
obtained by QCD sum rules~\cite{Grozin:1997pq}; no other estimate
exists for $\lambda_{E}$ or $\lambda_{H}$.
We calculate
(\ref{eq4}) 
for imaginary LC separation, performing the Wick rotation
$t\rightarrow -i\tau$~\cite{Grozin:1997pq,Braun:2003wx}.

%%%%%%%%%%%%%%%%%%%%%%%%%%%%%%%%%%%%%%%%%%%%%%%%%%%%%%%%%%%%%%%%%%%%%%%%%%%%%%%%%
\begin{figure}
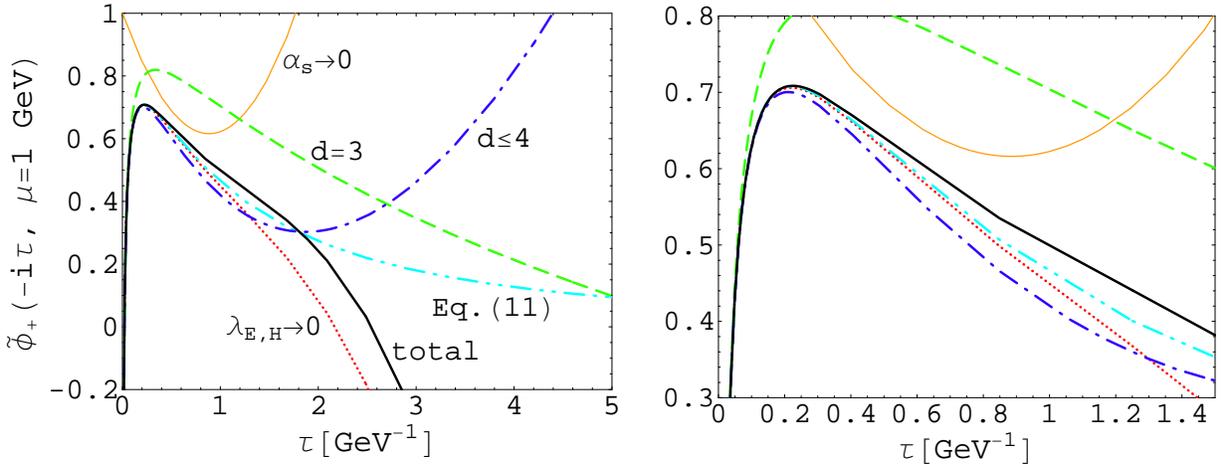

\begin{center}
\epsfig{figure=letterfig1.eps,width=0.50\textwidth,clip=}
\epsfig{figure=letterfig2.eps,width=0.49\textwidth,clip=}
\end{center}
\caption{The $B$-meson LCDA $\tilde{\phi}_+(-i\tau,\mu)$ at $\mu=1$~GeV
using the OPE-based form (\ref{eq4}):
%which includes all contributions
%from the local operators of dimension $d\leq 5$ and the corresponding Wilson coefficients
%to $O(\alpha_s)$:
%the ${\cal O}(\alpha_s)$ Wilson coefficients. 
the wide-solid curve represents the whole 
contributions, and 
the dashed and dot-dashed curves show the contributions associated with
the operators of dimension $d= 3$ and $d\leq 4$, respectively. 
The narrow-solid and dotted curves show the cases when $\alpha_s \rightarrow 0$
and $\lambda_{E,H}\rightarrow 0$, respectively.
The two-dot-dashed curve plots the result using 
the Lee-Neubert ansatz (\ref{LNmodel}).
}
\end{figure}
%%%%%%%%%%%%%%%%%%%%%%%%%%%%%%%%%%%%%%%%%%%%%%%%%%%%%%%%%%%%%%%%%%%%%%%%%%%%%%%%%%%%
%%%%%%%%%%%%%%%%%%%%%%%%%%%%%%%%%%%%%%%%%%%%%%%%%%%%%%%%%%%%%%%%%%%%%%%%%%%%%%%%%
%\begin{figure}
%\begin{center}
%\includegraphics[height=7cm]{letterfig2.eps}
%\end{center}
%\caption{Same as Fig.~1, within
%%showing the behaviors in 
%the small and moderate $\tau$ regions.
%% enlarged.
%%but for a smaller region of $\tau$.
%}
%%\label{fig:2}
%\end{figure}
%%%%%%%%%%%%%%%%%%%%%%%%%%%%%%%%%%%%%%%%%%%%%%%%%%%%%%%%%%%%%%%%%%%%%%%%%%%%%%%%%%%%
The results for $\tilde{\phi}_{+}(-i\tau, \mu=1~{\rm GeV})$ 
using (\ref{eq4}) are shown as
a function of $\tau$ in Fig.~1:
the RHS
shows the behaviors for
small and moderate $\tau$ regions,
and the LHS displays also the region with larger $\tau$.
The wide-solid curve 
shows the whole contributions 
of (\ref{eq4}),
while the narrow-solid curve shows the result
for $\alpha_s \rightarrow 0$; 
the NLO perturbative corrections are at the 10-30\% level for moderate $\tau$,
while they are very large for $\tau \rightarrow 0$ because of singular logarithms $L^2$ and $L$.
The dashed and dot-dashed curves show the contributions
of the first two terms and the first line in (\ref{eq4}), respectively,
associated with the operators of dimension 
$d =3$ and $d \le 4$,
while the dotted curve gives the results of (\ref{eq4}) 
when $\lambda_E=\lambda_H = 0$.
For moderate $\tau$, the contributions from the dimension-4 operators suppress the DA by
30-40\%, but
the dimension-5 operators, in contrast, lead to enhancement 
by 10-20\%
with significant effects from $\lambda_E$ and $\lambda_H$.

In the large $\tau$ region,
the dimension-4 and -5 operators 
change their roles such that they
enhance and suppress the DA, respectively,
because of the growth of
the double logarithm $L^2$ in the NLO Wilson coefficients;
this fact would eventually call for treating
the higher order perturbative corrections 
beyond NLO, in particular, resummation of the large logarithms in the 
Wilson coefficients.
We also note that the contributions from the dimension-$d$
operators
grow as $\sim \tau^{d-3}$
with increasing $\tau$,
and, beyond a certain value of $\tau$, 
the contributions from the dimension-4 or -5 operators become larger than those
from the lower-dimensional operators; this indicates that
the hierarchy of 
contributions is lost,
and
the OPE breaks down.
These considerations with 
quantitative results in Fig.~1
show that our $B$-meson LCDA (\ref{eq4})
indeed works up to moderate LC distances $\tau$
of order 1~GeV$^{-1} \sim 1/\mu$;
in this region, the hierarchy among the dashed, dot-dashed, and wide-solid curves
demonstrates convergence of the OPE (\ref{eq3}) and the corresponding accuracy 
of our LCDA (\ref{eq4}).

We find that the behavior of the wide-solid curve for small and moderate $\tau$ in Fig.~1 is
similar to that of the previous results~\cite{Braun:2003wx,Lee:2005gza} for the $B$-meson LCDA, 
obtained by taking into account perturbative as well as nonperturbative QCD corrections
in a systematic framework.
For example,
the two-dot-dashed curve in Fig.~1 shows
the behavior of the two-component ansatz
by Lee and Neubert~\cite{Lee:2005gza}, 
which is given in momentum space as
\begin{equation}
   \phi_+^{\rm LN}(\omega,\mu)
   = N\,\frac{\omega}{\omega_0^2}\,
    e^{-\omega/\omega_0} + \theta(\omega-\omega_t)\,
    \frac{C_F\alpha_s}{\pi\omega} 
 \left[ \left( \frac12 - \ln\frac{\omega}{\mu} \right)
    + \frac{4\bar\Lambda_{DA}}{3\omega}
    \left( 2 - \ln\frac{\omega}{\mu} \right) \right]\ ,
\label{LNmodel}
\end{equation}
where the second term reproduces the correct asymptotic behavior of the DA (\ref{eq1})
for $\omega \gg \Lambda_{\rm QCD}$,
and the first term represents the nonperturbative component 
modeled by an exponential form~\cite{Grozin:1997pq};
$\omega_t$ 
is chosen such that 
(\ref{LNmodel}) is continuous.
The other parameters, $N$ and $\omega_0$, are fixed by matching  
the first two ($j=0,1$) cut-moments 
$\int_0^{\Lambda_{UV}} d\omega  \omega^j 
\phi_+^{\rm LN} (\omega, \mu)$
with the OPE for the corresponding cut-moments $M_{0,1}$ derived in \cite{Lee:2005gza},
where the operators of dimension 
$d \le$ 4
and the corresponding Wilson coefficients at NLO are taken into account
(see the discussion below (\ref{eq4})).
The (central) values of these parameters are 
$\omega_t = 2.33$~GeV, $N=0.963$, and $\omega_0 = 0.438$~GeV at 
$\mu=1$~GeV~\cite{Lee:2005gza}.
For $\tau \lesssim 1$~GeV$^{-1}$ in Fig.~1, 
the Lee-Neubert ansatz (\ref{LNmodel}) shows behavior similar 
to (\ref{eq4}) with $\lambda_E=\lambda_H = 0$ substituted.
Indeed, when Fourier transformed to the coordinate space,
(\ref{LNmodel}) produces, with good accuracy, the terms in the first 
line of (\ref{eq4}), which are associated with the operators of dimension 
$d \le$ 4;
moreover, the first term of (\ref{LNmodel}) 
also produces particular
contributions associated with the operators of dimension-5 and higher 
(see (\ref{LNmodel1}) below),
and the sizes of those contributions are actually rather close
to those of the terms proportional to $\bar{\Lambda}^2$ 
in (\ref{eq4}) for $\tau \lesssim 1$~GeV$^{-1}$.

The wide-solid curve in Fig.~1 represents the model-independent 
behavior of the $B$-meson LCDA (\ref{eq1})
based on the most accurate OPE (\ref{eq3}), (\ref{eq4}) at present.
However, in the long-distance region, $\tau \gg 1$~GeV$^{-1}$,
the contributions associated with the operators of any higher dimension become important,
and the OPE diverges; thus, one has to rely on a certain model for 
the large $\tau$ behavior and connect the model-independent descriptions at small and moderate $\tau$
to that model in a reasonable manner.
The results in Fig.~1 suggest the possibility of connecting the behavior 
for $\tau \le \tau_c$ ($\tau_c \sim 1$~GeV$^{-1}$)
given by our OPE form  (\ref{eq4}) 
to that for $\tau \ge \tau_c$, given by the coordinate-space representation
of the first term of (\ref{LNmodel}),
\begin{equation}
\int_0^\infty d\omega e^{- \omega\tau}\ N\,\frac{\omega}{\omega_0^2}\,
    e^{-\omega/\omega_0}=
\frac{N}{\left( \tau \omega_0 +1\right)^2}\ .
\label{LNmodel1}
\end{equation}
Here, $N$ and $\omega_0$ 
can be determined such that
both the resulting total DA $\tilde{\phi}_+(-i\tau,\mu)$ and its derivative
$\partial \tilde{\phi}_+(-i\tau,\mu)/\partial \tau$ are continuous
at 
$\tau=\tau_c$.
Namely, we perform the matching of 
$\tilde{\phi}_+(-i\tau_c,\mu) = \int_0^\infty d\omega e^{-\omega\tau_c}\phi_+(\omega,\mu)$,
as well as of $\partial \tilde{\phi}_+(-i\tau_c,\mu)/\partial \tau_c
=- \int_0^\infty d\omega 
e^{-\omega\tau_c}\omega \phi_+(\omega,\mu)$,
between (\ref{eq4}) and
(\ref{LNmodel1}),
and
this is formally analogous to the matching used for (\ref{LNmodel}).
In the LHS of
Table~1, associated with the central values of (\ref{lambdaEH}), 
we show the values of $N$ and $\omega_0$ obtained by solving 
our matching relations for $\mu=1$~GeV. (The RHS of Table~1
shows 
the results that 
would be obtained by solving the similar matching 
relations 
with $\lambda_E=\lambda_H=0$,
and we find that, indeed, $\tau_c \simeq 0.7$~GeV$^{-1}$ gives the behavior
close to 
(\ref{LNmodel}).)
\begin{table}
\begin{center}
\begin{tabular}{|c|c|c|c||c|c|c|}
\hline
&\multicolumn{3}{|c||}{$\lambda_E^2=0.11$~{\small GeV$^2$},~~$\lambda_H^2=0.18$~{\small GeV$^2$}}
&
\multicolumn{3}{|c|}{$\lambda_E^2=\lambda_H^2=0$
}\\
\hline
$\tau_c$~{\small [GeV$^{-1}$]}& $N$ & $\omega_0$~{\small [GeV]} 
& $\lambda_B^{-1} 
$~{\small [GeV$^{-1}$]} 
& $N$ & $\omega_0$~{\small [GeV]} 
& $\lambda_B^{-1} 
$~{\small [GeV$^{-1}$]} \\
\hline
0.4  & 0.816 & 0.257 & $ 3.11\, \  ( 0.23  + 2.88 )$ 
     & 0.832 & 0.301 & $ 2.69\, \  ( 0.23  + 2.46 )$ \\
0.6  & 0.850 & 0.306 & $ 2.70\, \  ( 0.35  + 2.35 )$
     & 0.899 & 0.394 & $ 2.19\, \  ( 0.35 + 1.84 )$ \\
0.8  & 0.852 & 0.308 & $ 2.69\, \  ( 0.47  + 2.22 )$ 
     & 0.966 & 0.461 & $ 1.99\, \  ( 0.46 + 1.53 )$ \\
1.0  & 0.858 & 0.313 & $ 2.66\, \  ( 0.58  + 2.08  )$
     & 1.11 & 0.572 & $ 1.79\, \  ( 0.56 + 1.23 )$ \\
1.2  & 0.910 & 0.349 & $ 2.51\, \  ( 0.67  + 1.84 )$
     & 1.55 & 0.839 & $ 1.56\, \  ( 0.64 + 0.92 )$ \\
1.4  & 1.09  & 0.456 & $ 2.22\, \  ( 0.76  + 1.46 )$
     & 4.43 & 1.95 & $ 1.32\, \  ( 0.71 + 0.61 )$ \\
1.6  & 1.81  & 0.777 & $ 1.87\, \  ( 0.83  + 1.04 )$
     & 9.82 & $-4.55$ & $1.11\, \  ( 0.77 + 0.34 )$ \\
\hline
\end{tabular}
\end{center}
\caption{Parameters of the model function $(\ref{LNmodel1})$ for different values of $\tau_c$,
determined by continuity at $\tau=\tau_c$ 
with the OPE-based LCDA (\ref{eq4}) for $\mu=1$~GeV, and
the results of the inverse moment $\lambda_B^{-1}(\mu)$
at $\mu=1$~GeV, with the first and second 
numbers in the parentheses denoting the contributions from the first and the second terms
in the RHS of (\ref{lambdaB}).
}
\end{table}
In Fig.~2,
the wide-solid and two-dot-dashed curves
are same as those
in Fig.~1, 
and 
the dotted, 
solid-gray,
and dashed curves show the behavior of (\ref{LNmodel1}) 
for $\tau \ge \tau_c$ with 
$\tau_c=0.6$, $1.0$, and $1.4$~GeV$^{-1}$, 
respectively,
using the corresponding values of $N$ and $\omega_0$ in the LHS of
Table~1;
%%%%%%%%%%%%%%%%%%%%%%%%%%%%%%%%%%%%%%%%%%%%%%%%%%%%%%%%%%%%%%%%%%%%%%%%%%%%%%%%%
\begin{figure}[t!]
\begin{center}
\epsfig{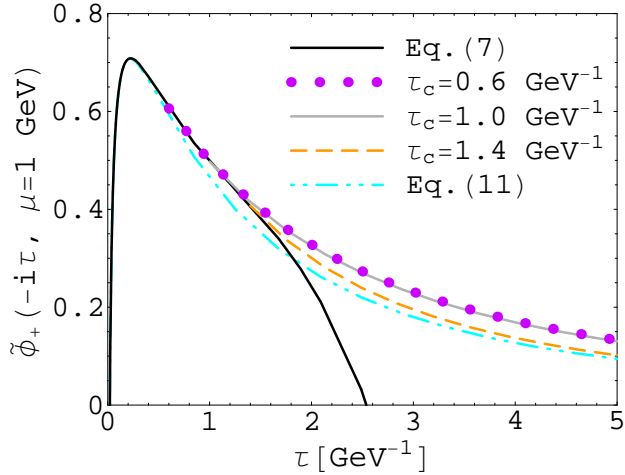}
\end{center}
\caption{The $B$-meson LCDA $\tilde{\phi}_+(-i\tau,\mu)$ at $\mu=1$~GeV.
The solid-black and two-dot-dashed curves show the results using
the OPE-based form (\ref{eq4}) and the Lee-Neubert ansatz (\ref{LNmodel}), respectively.   
The dotted, solid-gray and dashed curves plot the model function (\ref{LNmodel1}) 
for $\tau\geq\tau_c$ with $\tau_c=0.6,1.0$ and $1.4$~GeV$^{-1}$, respectively,
determined by continuity 
with (\ref{eq4}).
}
\end{figure}
%%%%%%%%%%%%%%%%%%%%%%%%%%%%%%%%%%%%%%%%%%%%%%%%%%%%%%%%%%%%%%%%%%%%%%%%%%%%%%%%%%%%
these three curves behave as 
$\sim N/(\omega_0^2 \tau^2)$ at large $\tau$,
with 
larger $N/\omega_0^2$ 
than those of
(\ref{LNmodel}) and the RHS of Table~1.
Indeed, 
we can show that
$N/\omega_0^2=(9/4 \bar{\Lambda}_{DA}^2)\left\{1+\tau_c \bar{\Lambda}_{DA}
\left[ \lambda_E^2/ \bar{\Lambda}_{DA}^2+\lambda_H^2/ (2\bar{\Lambda}_{DA}^2)
-1\right]\right\} +\cdots$, using our matching relations, 
and thus the contributions of 
$\lambda_E$ and $\lambda_H$ enhance $N/\omega_0^2$.

Using these results, 
we calculate the first inverse moment of the LCDA,
\begin{equation}
\lambda_B^{-1}(\mu) = \int_0^\infty d\omega
    \frac{\phi_+(\omega,\mu)}{\omega} = \int_0^{\tau_c} d\tau 
\tilde{\phi}_{+}(-i\tau, \mu)+ \int_{\tau_c}^\infty d\tau 
\tilde{\phi}_{+}(-i\tau, \mu)\ ,
\label{lambdaB}
\end{equation}
which is of particular interest for the QCD description of exclusive $B$-meson decays.
We substitute (\ref{eq4}) and (\ref{LNmodel1}) into the first and the second terms 
in the RHS, respectively, and the results 
are shown in Table~1
for each value of $\tau_c$.
The ``stable'' behavior observed for
$0.6~{\rm GeV}^{-1} \lesssim \tau_c \lesssim 1~{\rm GeV}^{-1}$ in the LHS of Table~1 
and in Fig.~2 suggests that $\lambda_B^{-1}(\mu=1~{\rm GeV})\simeq 2.7$~GeV$^{-1}$,
i.e., $\lambda_B(\mu=1~{\rm GeV})\simeq 0.37$~GeV.
This value of $\lambda_B$ is somewhat smaller 
than the previous estimates that include nonperturbative and/or perturbative
QCD corrections~\cite{Braun:2003wx,Lee:2005gza,Khodjamirian:2005ea}
(e.g., (\ref{LNmodel}) 
gives $\lambda_B(\mu=1~{\rm GeV}) \simeq 0.48$~GeV),
although consistent with them within their theoretical errors.
A value of $\lambda_B$ that is as small as our value 
was adopted in \cite{Beneke:2000ry,Beneke:2005vv}.
Note that in the RHS of Table~1 with $\lambda_{E,H} = 0$, 
the stable behavior is not seen as clearly as in the LHS,
and $\lambda_B$ assumes larger values than in the latter.

The above results demonstrate that 
the novel
HQET parameters, $\lambda_E$ and $\lambda_H$, 
associated with the dimension-5 quark-antiquark-gluon operators, 
could lead to smaller value  
of $\lambda_B$. 
In particular, this effect can be
more significant:
using 
the values 
of $\lambda_E$ and $\lambda_H$
corresponding to 
the upper bound of 
(\ref{lambdaEH}),
we find that 
the solid curve in Fig.~2 
becomes further
enhanced 
in the moderate $\tau$ region, so that (\ref{lambdaB}) gives
$\lambda_B(\mu=1~{\rm GeV})
\sim 0.2$~GeV or smaller.
This finding calls for 
more precise nonperturbative
estimates of 
$\lambda_E$ and $\lambda_H$.
We also note 
that in the RHS of (\ref{lambdaB}) evaluated in Table~1,
the second term is much larger than the first term.
This suggests that $\lambda_B$ is rather sensitive to 
the functional form
that models 
the LCDA in the long-distance region; for example,
a functional form motivated by $\phi^{(WW)}_+$, 
which is mentioned below (\ref{eq1}), 
provides an interesting possible 
alternative to (\ref{LNmodel1}).
Moreover, the Sudakov-type resummation 
for the Wilson coefficients
may give rise to additional important contribution in the large $\tau$ region.
Systematic investigations of these points, 
as well as the effect of the RG evolution
of the LCDA, are beyond the scope of this Letter and
will be presented elsewhere.

To summarize, 
we have derived the OPE 
that embodies both the notorious UV and IR behaviors of the $B$-meson LCDA, 
including all contributions from the local operators of dimension $d\leq 5$ and 
the corresponding Wilson coefficients at NLO accuracy.
This OPE 
allows us to 
parameterize all
nonperturbative contributions 
in terms of 
three HQET parameters
and provides us with 
the most accurate
description of the $B$-meson LCDA for distances less than $\sim 1/\Lambda_{\rm QCD}$. 
We have also used the model-independent behaviors from our OPE
to constrain the long-distance behavior of the LCDA and estimate 
the first inverse moment $\lambda_B^{-1}$ as the integral of the LCDA over 
entire distances. 
The results demonstrated the 
impact of the novel HQET parameters, associated with 
the matrix elements of the dimension-5 quark-antiquark-gluon 
operators.

\section*{Acknowledgments}
%\bigskip
We thank V.~M. Braun 
for valuable
discussions. 
This work is supported by the Grant-in-Aid for Scientific Research 
No.~B-19340063. The work of H.K. is supported in part by the UK Science \& 
Technology Facilities Council under grant number PP/E007414/1.


\begin{thebibliography}{99}
\bibitem{Beneke:2000ry}
  M.~Beneke, G.~Buchalla, M.~Neubert and C.~T.~Sachrajda,
  %``{QCD} factorization for B --> pi pi decays: Strong phases and CP  violation
  %in the heavy quark limit,''
  Phys.\ Rev.\ Lett.\  {\bf 83} (1999) 1914;
  Nucl. Phys. {\bf B591} (2000) 313; {\bf B606} (2001) 245.

\bibitem{Bauer:2001cu}
  C.~W.~Bauer, D.~Pirjol and I.~W.~Stewart,
  %``A proof of factorization for B --> D pi,''
  Phys.\ Rev.\ Lett.\  {\bf 87} (2001) 201806;
  %``Soft-collinear factorization in effective field theory,''
  %Phys. Rev. {\bf D65} (2002) 054022;
  %``Factorization and endpoint singularities in heavy-to-light decays,''
  Phys. Rev. {\bf  D67} (2003) 071502.\\
  C.~W.~Bauer, D.~Pirjol, I.~Z.~Rothstein and I.~W.~Stewart,
  %``B --> M(1) M(2): Factorization, charming penguins, strong phases, and
  %polarization,''
  Phys.\ Rev.\ {\bf D70} (2004) 054015.
%     [arXiv:hep-ph/0401188].  

\bibitem{Li:2003yj}
 H.~n.~Li and H.~L.~Yu,
%  %``Extraction of V(ub) from decay B $\to$ pi lepton neutrino,''
  Phys.\ Rev.\ Lett.\  {\bf 74} (1995) 4388;
%  %``Pqcd Analysis Of Exclusive Charmless B Meson Decay Spectra,''
  Phys.\ Lett.\ {\bf B353} (1995) 301;
%  %``Perturbative QCD Analysis Of B Meson Decays,''
 Phys.\ Rev.\ {\bf D53} (1996) 2480.
 % H.~n.~Li,
 % %``QCD aspects of exclusive B meson decays,''
 % Prog.\ Part.\ Nucl.\ Phys.\  {\bf 51} (2003) 85. 
  %and references therein.

\bibitem{Szczepaniak:1990dt}
  A.~Szczepaniak, E.~M.~Henley and S.~J.~Brodsky,
%  %``PERTURBATIVE QCD EFFECTS IN HEAVY MESON DECAYS,''
  Phys.\ Lett.\  B {\bf 243} (1990) 287.
%\bibitem{Akhoury:1993uw}
%  R.~Akhoury, G.~Sterman and Y.~P.~Yao,
%  %``Exclusive semileptonic decays of B mesons into light mesons,''
%  Phys.\ Rev.\ {\bf D50} (1994) 358.
%\bibitem{Korchemsky:1999qb}

\bibitem{Beneke:2000wa}
  G.~P.~Korchemsky, D.~Pirjol and T.~M.~Yan,
  %``Radiative leptonic decays of B mesons in QCD,''
  Phys.\ Rev.\ {\bf D61} (2000) 114510.\\
  M.~Beneke and T.~Feldmann,
  %``Symmetry-breaking corrections to heavy-to-light B meson form factors at
  %large recoil,''
  Nucl.\ Phys.\ {\bf B592} (2001) 3.\\
 % M.~Beneke, A.~P.~Chapovsky, M.~Diehl and T.~Feldmann,
 % %``Soft-collinear effective theory and heavy-to-light currents beyond  leading
 % %power,''
 % Nucl.\ Phys.\ {\bf B643} (2002) 431.\\
  %    [arXiv:hep-ph/0008255].
  S.~W.~Bosch and G.~Buchalla,
  %``The radiative decays B --> V gamma at next-to-leading order in QCD,''
  Nucl.\ Phys.\ {\bf B621} (2002) 459;
  %``The double radiative decays B --> gamma gamma in the heavy quark limit.
  %((U)),''
  JHEP {\bf 0208} (2002) 054.\\
%D.~Pirjol and I.~W.~Stewart,
 % %``A complete basis for power suppressed collinear-ultrasoft operators,''
 % Phys.\ Rev.\ {\bf D67} (2003) 094005
 % [Erratum-ibid.\ {\bf D69} (2004) 019903].\\
  B.~Grinstein and D.~Pirjol,
  %``The forward-backward asymmetry in B --> K pi l+ l- decays,''
  Phys.\ Rev.\ {\bf  D73} (2006) 094027; {\bf  D73} (2006) 014013.\\
 C.~W.~Bauer, I.~Z.~Rothstein and I.~W.~Stewart,
  %``SCET analysis of B --> K pi, B --> K anti-K, and B --> pi pi decays,''
  Phys.\ Rev.\ {\bf D74} (2006) 034010.\\
%\bibitem{KLS} 
 Y.~Y.~Keum, H.~n.~Li and A.~I.~Sanda,
  %``Fat penguins and imaginary penguins in perturbative QCD,''
  Phys.\ Lett.\ {\bf B504} (2001) 6;
  %``Penguin enhancement and B --> K pi decays in perturbative QCD,''
  Phys.\ Rev.\ {\bf  D63} (2001) 054008.\\ 
  T.~Kurimoto, H.~n.~Li and A.~I.~Sanda,
  %``Leading power contributions to B --> pi, rho transition form factors,''
  Phys.\ Rev.\ {\bf  D65} (2002) 014007.

\bibitem{BBNL}
  S.~Descotes-Genon and C.~T.~Sachrajda,
  %``Factorization, the light-cone distribution amplitude of the B-meson and the
  %radiative decay B --> gamma l nu/l. ((V)),''
  Nucl.\ Phys.\ {\bf  B650} (2003) 356;
  %``Universality of nonperturbative QCD effects in radiative B decays. ((U)),''
  Phys.\ Lett.\ {\bf B557} (2003) 213.\\
  S.~W.~Bosch, R.~J.~Hill, B.~O.~Lange and M.~Neubert,
  %``Factorization and Sudakov resummation in leptonic radiative B decay,''
  Phys.\ Rev.\ {\bf D67} (2003) 094014.\\
  E.~Lunghi, D.~Pirjol and D.~Wyler,
  %``Factorization in leptonic radiative B --> gamma e nu decays. ((U)),''
   Nucl.\ Phys.\ {\bf B649} (2003) 349.

%\cite{Beneke:2005vv}
\bibitem{Beneke:2005vv}
  M.~Beneke and M.~Neubert,
  %``QCD factorization for B --> P P and B --> P V decays,''
  Nucl.\ Phys.\ {\bf B675} (2003) 333.\\
  M.~Beneke and S.~Jager,
  %``Spectator scattering at NLO in non-leptonic B decays: Tree amplitudes,''
  Nucl.\ Phys.\ {\bf B751} (2006) 160;
%    M.~Beneke and S.~Jager,
  %``Spectator scattering at NLO in non-leptonic B decays: Leading penguin
  %amplitudes,''
%  Nucl.\ Phys.\ 
  {\bf  B768} (2007) 51.

\bibitem{Bell08}
  N.~Kivel,
  %``Radiative corrections to hard spectator scattering in $B\to \pi\pi$
  %decays,''
  JHEP {\bf 0705} (2007) 019.\\
  V.~Pilipp,
  %``Hard spectator interactions in B to pi pi at order alpha_s^2,''
  Nucl.\ Phys.\ {\bf B794} (2008) 154.\\
  G.~Bell,
  %``NNLO Vertex Corrections in charmless hadronic B decays: Imaginary part,''
  Nucl.\ Phys.\ {\bf  B795} (2008) 1.




\bibitem{Grozin:1997pq}
A.~G.~Grozin and M.~Neubert,
Phys. Rev. {\bf D55} (1997) 272.
%[hep-ph/9607366].


\bibitem{Ball:2008fw}
  P.~Ball, V.~M.~Braun and E.~Gardi,
  %``Distribution Amplitudes of the Lambda_b Baryon in QCD,''
  Phys.\ Lett.\ {\bf B665} (2008) 197.

\bibitem{BL}
  G.~P.~Lepage and S.~J.~Brodsky,
  %``Exclusive Processes In Perturbative Quantum Chromodynamics,''
  Phys.\ Rev.\ {\bf D22} (1980) 2157.
%V.~L.~Chernyak and A.~R.~Zhitnitsky,
%Phys. Rept. {\bf 112} (1984) 173.\\
%  S.~J.~Brodsky and G.~P.~Lepage,
%  %``Exclusive Processes In Quantum Chromodynamics,''
%  Adv.\ Ser.\ Direct.\ High Energy Phys.\  {\bf 5} (1989) 93.


\bibitem{Neubert:1994mb}
M.~Neubert,
Phys. Rept. {\bf 245} (1994) 259.
%, and references therein.
%[hep-ph/9306320].


\bibitem{KKQT}
H.~Kawamura, J.~Kodaira, C.F.~Qiao and K.~Tanaka,
Phys. Lett. {\bf B523} (2001) 111;
Erratum-ibid. {\bf B536} (2002) 344; 
Mod. Phys. Lett. {\bf A18} (2003) 799.
%Nucl. Phys. B (Proc. Suppl.) {\bf 116} (2003) 269.

%  T.~Huang, X.~G.~Wu and M.~Z.~Zhou,
%  %``B-meson wavefunction in the Wandzura-Wilczek approximation,''
%  Phys.\ Lett.\ {\bf  B611} (2005) 260.

\bibitem{Huang:2005kk}
%\bibitem{Khodjamirian:2006st}
  T.~Huang, C.~F.~Qiao and X.~G.~Wu,
  %``B-meson wavefunction with 3-particle Fock states' contributions,''
  Phys.\ Rev.\ {\bf D73} (2006) 074004.\\
  B.~Geyer and O.~Witzel,
  %``Heavy Meson Distribution Amplitudes of Definite Geometric Twist with
  %Contribution of 3-Particle Distribution Amplitudes,''
  Phys.\ Rev.\ {\bf  D72} (2005) 034023; {\bf  D76} (2007) 074022.\\
  A.~Khodjamirian, T.~Mannel and N.~Offen,
  %``Form factors from light-cone sum rules with B-meson distribution
  %amplitudes,''
  Phys.\ Rev.\ {\bf D75} (2007) 054013.


\bibitem{Lange:2003ff}
B.~O.~Lange and M.~Neubert,
  %``Renormalization-group evolution of the B-meson light-cone distribution
  %amplitude,''
Phys. Rev. Lett.  {\bf 91} (2003) 102001.

\bibitem{Korchemsky:1987wg}
%G.~P.~Korchemsky and A.~V.~Radyushkin,
%%``Renormalization of the Wilson Loops Beyond the Leading Order,''
%Nucl.\ Phys.\ {\bf B283} (1987) 342.\\
%　　　　\bibitem{Korchemskaya:1992je}
I.~A.~Korchemskaya and G.~P.~Korchemsky,
%``On lightlike Wilson loops,''
Phys.\ Lett.\ {\bf  B287} (1992) 169.

\bibitem{Braun:2003wx}
V.~M.~Braun, D.~Y.~Ivanov and G.~P.~Korchemsky,
  %``The B-meson distribution amplitude in QCD,''
Phys. Rev. {\bf D69} (2004) 034014.

\bibitem{Lee:2005gza}
S.~J.~Lee and M.~Neubert,
  %``Model-independent properties of the B-meson distribution amplitude,''
Phys. Rev. {\bf D72} (2005) 094028.

\bibitem{Abbott:1980hw}
J.~S.~Schwinger,
  %``On gauge invariance and vacuum polarization,''
Phys. Rev. {\bf 82} (1951) 664.\\
E.~V.~Shuryak and A.~I.~Vainshtein,
Nucl. Phys. 
  %``Theory Of Power Corrections To Deep Inelastic Scattering In Quantum
  %Chromodynamics. 1. Q**2 Effects,''
%Nucl.\ Phys.\  B 
%ibid. 
{\bf B199} (1982) 451; {\bf B201} (1982) 141.\\
%V.~A.~Novikov, M.~A.~Shifman, A.~I.~Vainshtein and V.~I.~Zakharov,
%  %``Calculations In External Fields In Quantum Chromodynamics. Technical
%  %Review,''
%Fortsch.\ Phys.\  {\bf 32} (1984) 585.\\
I.~I.~Balitsky and V.~M.~Braun,
  %``Evolution Equations for QCD String Operators,''
%  Nucl.\ Phys.\  B 
Nucl. Phys. {\bf B311} (1989) 541.


\bibitem{KT08} 
H.~Kawamura and K.~Tanaka, in preparation.

\bibitem{GN97}
  A.~G.~Grozin and M.~Neubert,
  %``Hybrid renormalization of penguins and dimension-5 heavy-light
  %operators,''
  Nucl.\ Phys.\ {\bf B495} (1997) 81.

%\bibitem{Bauer:2003piBauer:2003pi}
%  C.~W.~Bauer and A.~V.~Manohar,
%  %``Shape function effects in B --> X/s gamma and B --> X/u l nu decays,''
%  Phys.\ Rev.\ {\bf  D70} (2004) 034024.

%\bibitem{KS94}
%  G.~P.~Korchemsky and G.~Sterman,
%  %``Infrared factorization in inclusive B meson decays,''
%  Phys.\ Lett.\ {\bf B340} (1994) 96.\\
%  A.~G.~Grozin and G.~P.~Korchemsky,
%  %``Renormalized sum rules for structure functions of heavy mesons decays,''
%  Phys.\ Rev.\ {\bf D53} (1996) 1378.


\bibitem{Bell:2008er}
  G.~Bell and T.~Feldmann,
  %``Modelling light-cone distribution amplitudes from non-relativistic bound
  %states,''
  JHEP {\bf 0804} (2008) 061.

\bibitem{Braun:1990iv}
V.~M.~Braun and I.~E.~Filyanov,
Z. Phys. {\bf C48} (1990) 239.\\
P.~Ball,
% et al.,
V.~M.~Braun, Y.~Koike and K.~Tanaka,
Nucl. Phys. {\bf B529} (1998) 323.


\bibitem{Bigi:1994}
I.~I.~Y.~Bigi, M.~A.~Shifman, N.~G.~Uraltsev and A.~I.~Vainshtein,
Phys.\ Rev.\ {\bf D50} (1994) 2234.\\
M.~Beneke and V.~M.~Braun,
Nucl.\ Phys.\ {\bf B426} (1994) 301.

\bibitem{Khodjamirian:2005ea}
  P.~Ball and E.~Kou,
  %``B --> gamma e nu transitions from QCD sum rules on the light-cone,''
  JHEP {\bf 0304} (2003) 029.\\
  A.~Khodjamirian, T.~Mannel and N.~Offen,
  %``B-meson distribution amplitude from the B --> pi form factor,''
  Phys.\ Lett.\ {\bf B620} (2005) 52.
\end{thebibliography}
\end{document}